\documentclass{article}

\usepackage{PRIMEarxiv}

\usepackage[utf8]{inputenc} % allow utf-8 input
\usepackage[T1]{fontenc}    % use 8-bit T1 fonts
\usepackage{hyperref}       % hyperlinks
\usepackage{url}            % simple URL typesetting
\usepackage{booktabs}       % professional-quality tables
\usepackage{amsfonts}       % blackboard math symbols
\usepackage{nicefrac}       % compact symbols for 1/2, etc.
\usepackage{microtype}      % microtypography
\usepackage{lipsum}
\usepackage{graphicx}
\usepackage{subcaption} 
\usepackage{pgfplots} 
\usepackage{xcolor} 
\usepackage{listings} 
\usepackage{tcolorbox}  
\usepackage{pgfplots}
\usepackage{float}
\usepackage{etoolbox}
\usepackage{amsmath,amsfonts}
\usepackage{algorithm}
\usepackage{algorithmic}
\usepackage{array}
\usepackage{textcomp}
\usepackage{stfloats}
\usepackage{url}
\usepackage{verbatim}
\usepackage{graphicx}
\usepackage{enumerate}
\usepackage[T1]{fontenc}
\usepackage{multirow}
\usepackage{makecell}
\graphicspath{{media/}}     % organize your images and other figures under media/ folder

\definecolor{codeblue}{RGB}{41,123,182}    % 更柔和的蓝色用于关键字
\definecolor{codegreen}{RGB}{80,161,79}    % 更柔和的绿色用于注释
\definecolor{codepurple}{RGB}{157,97,184}  % 紫色用于类型
\definecolor{codeorange}{RGB}{220,130,60}  % 橙色用于字符串
\definecolor{codegray}{RGB}{145,145,145}   % 灰色用于行号
\definecolor{backcolour}{RGB}{250,250,250} % 非常浅的灰色背景

\title{Distilling Lightweight Language Models for \\ C/C++ Vulnerabilities
}

\author{
  Zhiyuan Wei\thanks{These authors contributed equally to this work.} \\
  School of Computer Science and Technology \\
  Beijing Institute of Technology \\
  Beijing, China\\
  \texttt{weizhiyuan@bit.edu.cn} \\
  \And
  Xiaoxuan Yang\footnotemark[1] \\
  School of Cyberspace Science and Technology \\
  Beijing Institute of Technology \\
  Beijing, China\\
  \texttt{yangxiaoxuan123@bit.edu.cn} \\
  \AND
  Jing Sun \\
  School of Computer Science \\
  University of Auckland \\
  Auckland, New Zealand\\
  \texttt{jing.sun@auckland.ac.nz} \\
  \And
  Zijian Zhang \\
  School of Cyberspace Science and Technology \\
  Beijing Institute of Technology \\
  Beijing, China\\
  \texttt{zhangzijian@bit.edu.cn} \\
}

\begin{document}
\maketitle

\begin{abstract}
The increasing complexity of modern software systems exacerbates the prevalence of security vulnerabilities, posing risks of severe breaches and substantial economic loss. Consequently, robust code vulnerability detection is essential for software security. While Large Language Models (LLMs) have demonstrated remarkable capabilities in natural language processing, their potential for automated code vulnerability detection remains underexplored. This paper presents FineSec, a novel framework that harnesses LLMs through knowledge distillation to enable efficient and precise vulnerability identification in C/C++ codebases. FineSec utilizes knowledge distillation to transfer expertise from large teacher models to compact student models, achieving high accuracy with minimal computational cost. By integrating data preparation, training, evaluation, and continuous learning into a unified, single-task workflow, FineSec offers a streamlined approach. Extensive evaluations on C/C++ codebases demonstrate its superiority over both base models and larger LLMs in identifying complex vulnerabilities and logical flaws, establishing FineSec as a practical and scalable solution for real-world software security. To facilitate reproducibility, the datasets, source code, and experimental results are made publicly available at: \url{https://github.com/yangxiaoxuan123/FineSec_detect}.
\end{abstract}

% keywords can be removed

\section{Introduction}
The rapid growth in software complexity has led to a surge in security vulnerabilities, which, if exploited, can result in severe breaches and substantial economic losses. Effective code vulnerability detection is therefore essential for ensuring software security and reliability. As of July 2025, the National Vulnerability Database (NVD) reported 26,232 new vulnerabilities \cite{1}, marking a record high. This sharp increase underscores the escalating complexity of modern software systems and the persistent emergence of new security threats. The proportion of ultra-high-risk vulnerabilities continues to rise, further intensifying the urgency of robust detection measures.

Among programming languages, C/C++ remains a cornerstone in domains such as operating systems, embedded systems, and large-scale application development. However, its foundational design exposes it to unique security risks. For instance, C lacks automatic bounds checking for arrays and strings, allowing out-of-bounds memory access that can lead to buffer overflows \cite{2}. In March 2023, OpenSSL was found to contain a critical buffer overflow vulnerability (CVE-2023-4911) \cite{3}. Similarly, the Glibc library was discovered to have a buffer overflow flaw in environment variable handling, enabling local privilege escalation and bypassing sandbox protections across all major Linux distributions, including Ubuntu Snap and Flatpak. This vulnerability was ranked among the top three most dangerous of 2023 by MITRE, with significant security and financial impacts. These examples highlight the critical and urgent need for advanced vulnerability detection techniques, particularly for C-based software, to safeguard system security and reliability.

Source code vulnerability detection techniques have advanced considerably, with common approaches including symbolic execution–based analysis and fuzz testing. While effective in many scenarios, these methods face practical limitations. Fuzz testing (e.g., AFL) \cite{4} typically requires compiling source code and struggles to detect vulnerabilities in complex systems. Similarly, symbolic execution–based techniques often depend on generating LLVM intermediate representations (LLVM bitcode) \cite{5}, making their application reliant on compilation and bitcode generation processes. Traditional machine learning–based solutions have improved detection efficiency but remain constrained by their limited adaptability, as they are usually tailored to specific programming languages or vulnerability categories.

Large Language Models (LLMs), such as OpenAI’s GPT series \cite{6}, excel at understanding and generating text, making them well-suited for static code analysis, where source code can be treated as a specialized form of text. By learning both structural and semantic patterns, LLMs can detect syntax errors, violations of coding conventions, and potential security flaws. Bug detection reports share similarities with natural language processing tasks such as text summarization, where ChatGPT and other LLMs have demonstrated exceptional performance \cite{7}. Recent studies have explored applying LLMs to vulnerability detection by leveraging large datasets and diverse training strategies. However, their effectiveness across different vulnerability types, datasets, and fine-tuning approaches remains insufficiently studied. Nvidia\cite{belcak2025smalllanguagemodelsfuture} has put forward a forward-looking viewpoint: in the field of agents, small models will eventually replace large models. This is because Agent applications (such as autonomous decision-making and tool invocation) require extremely high real-time performance, cost, and deployment flexibility, and small models, with their characteristics of small size, fast response, and easy fine-tuning, can better meet these requirements. How to balance and even improve accuracy while ensuring high efficiency with small models has become a key issue that urgently needs to be addressed. In this work, we introduce knowledge extraction and model enhancement techniques to develop specialized lightweight LLM-based models for detecting vulnerabilities in C/C++ language source code. Our main contributions are:

\begin{itemize}
\item \textbf{Automatic Framework:} We present \textit{FineSec}, an automated framework that integrates data preprocessing, knowledge distillation, parameter-efficient fine-tuning via QLoRA, and continual learning. FineSec enables efficient, scalable adaptation of LLMs for robust and generalizable vulnerability detection.

\item \textbf{Domain-Specific LLMs:} FineSec serves as a pre-training framework tailored to vulnerability detection in C/C++ code, substantially improving detection accuracy while retaining adaptability to other programming languages.

\item \textbf{Evaluation and Benchmarking:} We benchmark seven representative LLMs, before and after FineSec fine-tuning, on synthetic and real-world datasets spanning 30+ CWE categories. Comparisons with other base models reveal how architecture, parameter scale, dataset type, and fine-tuning strategy influence detection performance.

\item \textbf{New Vulnerability Discovery:} FineSec uncovers vulnerabilities not labeled in existing CWE datasets. Manual inspection identified more than nine previously undocumented vulnerability patterns in C/C++ code, including subtle logic flaws and overlooked memory manipulation edge cases, demonstrating strong generalization and proactive detection capabilities.

\item \textbf{Fine-grained Vulnerability Analysis Framework:} We propose a fine-grained vulnerability analysis framework for vulnerability detection based on LLMs - categorizing prediction errors into five major types (memory management, input validation, information security, permission control, and concurrent problems), and analyze prediction errors to identify key bottlenecks in LLM-based vulnerability detection, offering guidance for building more robust and efficient detection systems.
\end{itemize}

The remainder of this paper is organized as follows. Section 2 provides background on C/C++ language vulnerabilities, the application of LLMs in software security, and knowledge distillation. Section 3 introduces the FineSec framework, detailing the knowledge generation process and the three-stage training pipeline. Section 4 outlines the research questions, describes the experimental design, and analyzes the results. Section 5 reviews related work on LLM-based vulnerability prediction, discusses the findings, and examines potential threats to validity. Finally, Section 6 concludes the paper and highlights future research directions for applying LLMs to program vulnerability detection.

\section{Background}
\subsection{Security Consideration for C/C++ Programming Languages}
The C/C++ programming language was designed to provide portable and efficient access to hardware resources while maintaining minimal abstraction over the underlying machine architecture. This design enables direct memory manipulation, fine-grained control over data representation, and deterministic execution—features that are critical for operating systems, embedded systems, and performance-sensitive applications. However, this philosophy also introduces inherent characteristics that significantly increase susceptibility to security risks. Unlike modern memory-safe languages (e.g., Rust, Go), C/C++ lacks built-in safeguards against common programming errors, relying instead on developer discipline to prevent vulnerabilities. As a result, decades of critical software exploits, from buffer overflows to privilege escalations, can be traced to the permissive memory model of C / C++, unchecked pointer arithmetic and legacy unsafe functions.

Several programming features of C/C++ contribute directly to its vulnerability profile. Firstly, unsafe memory management \cite{8} arises from the use of functions such as malloc and free, which provide developers full control over memory management without automatic garbage collection or built-in boundary/type tracking for allocated memory blocks. This often results in memory overflows and leaks (e.g., CWE-787, CWE-125). These vulnerabilities may allow attackers to execute arbitrary code by overwriting critical memory structures, bypassinging security mechanisms through memory corruption, or crash systems via resource exhaustion. Secondly, improper use of pointers \cite{9} remains a significant security risk in both C and C++. While C++ offers smart pointers to mitigate some issues, its raw pointers—like those in C—still lack bounds checking, risking invalid memory access and undefined behavior (e.g., CWE-119). The absence of lifetime tracking makes null pointer dereferences common (e.g., CWE-476), and improper validation when accessing null pointers can lead to system crashes. Pointers also permit arbitrary type casting of memory, creating opportunities for control-flow hijacking and data leakage. Lastly, format string functions \cite{10}, such as printf and sprintf, are widely used in C/C++, but improper use can lead to format string vulnerabilities (e.g. CWE-20). If user input is directly used as a format string, special specifiers like \%n can write to arbitrary memory. The lack of compile-time checking allows attackers to exploit these vulnerabilities to execute arbitrary code or leak sensitive information.

The number of vulnerabilities in C/C++ is large and is continuously growing. The Common Weakness Enumeration (CWE) system classifies common security flaws across programming languages, defining each with a specific and concise description. The official CWE website \cite{11} has published the \textit{Top 10 Most Dangerous Software Weaknesses of 2023}, including categories such as Input Validation \& Injection, Memory Safety, and System Resource \& Concurrency. These vulnerabilities are among the most common and critical in software today, enabling attackers to take full control of systems, steal sensitive data, or disrupt application functionality. They highlight the inherent risks of C/C++’s design and, due to their prevalence and severity, remain highly relevant in real-world security research. In Section~4, we provide a deeper analysis of these CWEs.

\subsection{LLMs for Software Security}

Modern vulnerability detection methods primarily fall into three categories: static analysis, machine learning, and LLMs. Static analysis detects vulnerabilities through predefined rules, pattern matching, and code flow analyses(e.g., CodeQL, FlawFinder, CPPCheck \cite{16,17,18}), while machine learning approaches leverage neural architectures to identify patterns from code representations like sequences or graphs\cite{20,21,22}. Nowadays, the rise of LLM has had a significant impact on the prospects of vulnerability detection: LLM was originally developed for natural language processing (NLP) tasks such as text generation and machine translation \cite{12}, and since code can be seen as a special type of language, there are now many studies utilizing LLM to process code - vulnerability detection is a critical application that can maintain system security and ensure asset integrity.

Trained on large-scale code repositories, LLMs can learn latent patterns—particularly those associated with security vulnerabilities—and reason about code logic in context. This enables more effective detection, such as tracing potential vulnerability paths through function calls, memory allocations, and pointer operations. The emergence of ChatGPT has accelerated advances in LLM-based vulnerability detection. Gao et al. \cite{24} reported that GPT-4 significantly outperformed both traditional deep learning models and static analyzers in a benchmark study evaluating 16 LLMs for C language vulnerability detection. Wei et al. \cite{10.1145/3695864} conducted a comprehensive evaluation of smart contract automation analysis tools, conducting experiments on 14 traditional detection tools and large models that meet the standards in terms of detection applicability, resource efficiency, version compatibility, and category coverage. The results showed that GPT-4o performed the best in most cases, followed by Llama-3.1-8b, reflecting the comprehensive vulnerability detection capability of the large model. Khare et al. \cite{39} showed that optimized prompting strategies can further boost LLM detection performance, while fine-tuned models such as CodeT5 and NatGen achieve substantial improvements on vulnerability detection tasks. These findings highlight LLMs’ potential to advance automated vulnerability detection and open new research directions in software security.

\subsection{Knowledge Distillation}
\label{subsec:knowledge_distillation}
Knowledge distillation (KD) \cite{28} for large-scale language models has emerged as a key technique for transferring knowledge from powerful teacher models to more efficient student models. The process typically comprises two core stages: knowledge elicitation and distillation algorithms.
The knowledge elicitation stage focuses on extracting valuable information from teacher models through various strategies. These include labeling (generating task-specific outputs for seed data), expansion (leveraging in-context learning to create similar data), synthesis (producing data from scratch using metadata), feature extraction (capturing internal representations from open-source teacher models), feedback (generating assessments or corrections), and self-knowledge (allowing student models to iteratively refine their outputs). These strategies are often combined to improve overall effectiveness, with the goal of acquiring specific skills, domain expertise, or general capabilities from the teacher model.

In the distillation stage, the extracted knowledge is injected into student models using methods such as supervised fine-tuning (maximizing the likelihood of teacher outputs), divergence and similarity alignment (matching probability distributions or hidden states), reinforcement learning from artificial intelligence feedback (RLAIF), and ranking-based optimization approaches such as Direct Preference Optimization (DPO) and Rank Responses to Align Language Models with Human Feedback(RRHF)\cite{29}. Based on the nature of the transferred capabilities, KD can be categorized into skill distillation and vertical domain distillation. Skill distillation focuses on general abilities such as instruction following, alignment, and multimodal processing. Vertical domain distillation targets specialized domains such as law, healthcare, finance, and science, and can equally be applied to vulnerability detection to enhance the expertise of student models in security knowledge. Both rely on the interplay between knowledge elicitation and distillation algorithms to effectively adapt LLMs to specific tasks or domains.

\section{Methodology}
\label{sec:methodology}
This section presents the FineSec framework, a comprehensive approach to vulnerability detection that leverages knowledge distillation to transfer domain-specific expertise from large teacher models to efficient student models.

\subsection{Framework Overview}
\label{subsec:framework_overview}

\subsubsection{Problem Definition}
\label{subsec:problem_definition}

The vulnerability detection problem can be formally defined as a binary classification task over source code snippets. Given a code snippet $x \in \mathcal{X}$, where $\mathcal{X}$ represents the space of all possible code fragments in a target programming language (C/C++ in our case), the objective is to learn a function $f: \mathcal{X} \rightarrow \{0, 1\}$ that accurately predicts whether $x$ contains a security vulnerability.

More specifically, let $\mathcal{D} = \{(x_i, y_i)\}_{i=1}^N$ be a dataset of code snippets where $x_i \in \mathcal{X}$ and $y_i \in \{0, 1\}$ indicates the presence (1) or absence (0) of vulnerabilities. The goal is to learn a model $f_\theta$ parameterized by $\theta$ that minimizes the expected risk:

\begin{equation}
\mathcal{R}(f_\theta) = \mathbb{E}_{(x,y) \sim \mathcal{P}} [\ell(f_\theta(x), y)]
\end{equation}

where $\mathcal{P}$ is the true data distribution and $\ell$ is a loss function (typically cross-entropy for classification tasks).

However, the challenge in vulnerability detection extends beyond simple binary classification. Real-world applications require models that can not only detect vulnerabilities but also provide detailed explanations, classify vulnerability types according to established taxonomies (such as CWE), and suggest appropriate remediation strategies. Therefore, we extend the problem formulation to include structured output generation:

\begin{equation}
f_\theta: \mathcal{X} \rightarrow \mathcal{Y}
\end{equation}

where $\mathcal{Y}$ represents the space of structured vulnerability reports containing vulnerability classification, root causes, severity assessment, location information, and remediation suggestions.

\subsubsection{Knowledge Distillation Framework}
The core innovation of FineSec lies in its application of knowledge distillation to transfer vulnerability detection expertise from large, computationally expensive teacher models to smaller, more efficient student models. This approach addresses the practical constraints of deploying large language models in resource-limited environments while maintaining high detection accuracy.

Let $T$ denote a large teacher model (such as GPT-4o) with parameters $\theta_T$, and $S$ denote a smaller student model with parameters $\theta_S$. The knowledge distillation process aims to train the student model to approximate the teacher's behavior on vulnerability detection tasks:

\begin{equation}
\mathcal{L}_{KD} = \alpha \mathcal{L}_{CE}(S(x), y) + (1-\alpha) \mathcal{L}_{KL}(S(x), T(x))
\end{equation}

where $\mathcal{L}_{CE}$ is the cross-entropy loss with ground truth labels, $\mathcal{L}_{KL}$ is the Kullback-Leibler divergence between student and teacher outputs, and $\alpha$ is a weighting parameter that balances the two objectives.

\subsubsection{FineSec Architecture}

\begin{figure}[t]
    \centering
    \includegraphics[width=0.95\linewidth, height=0.8\linewidth, keepaspectratio]{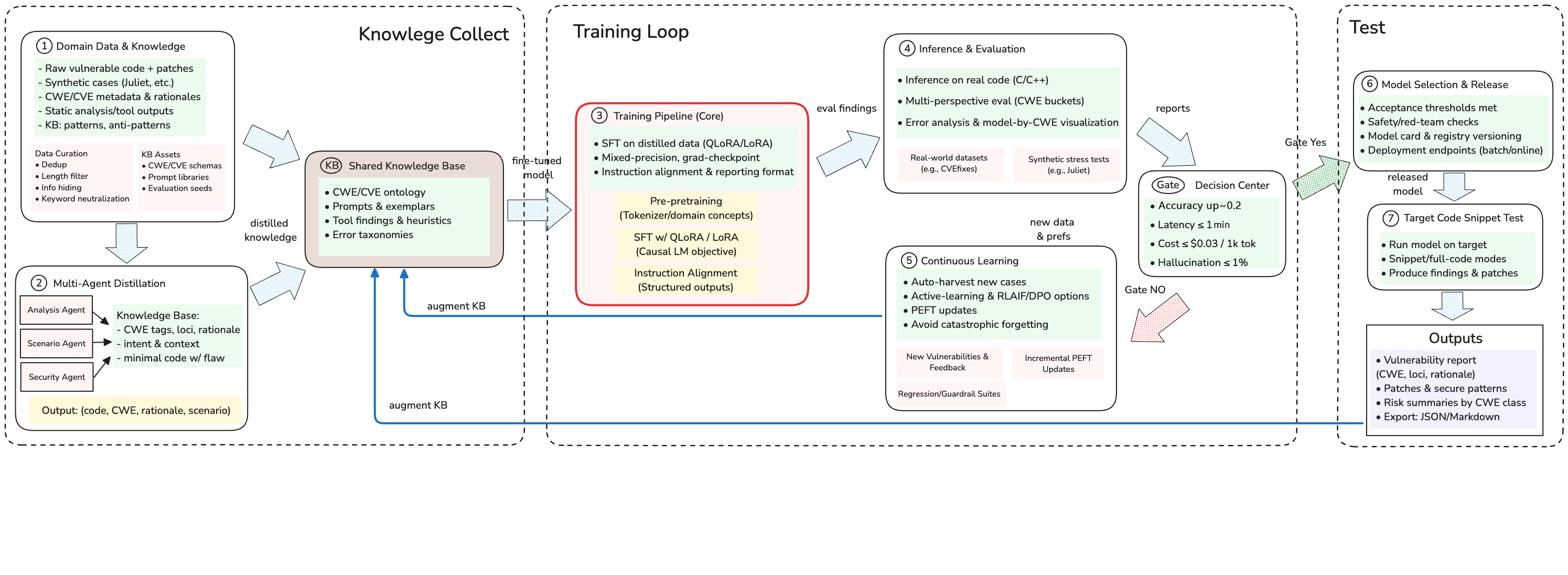}
    \caption{An Overview of FineSec Framework}
    \label{fig:overall}
\end{figure}

The FineSec framework consists of eight interconnected components organized in a pipeline that transforms raw vulnerability data into trained, deployable models capable of accurate vulnerability detection. Figure~\ref{fig:overall} illustrates the complete architecture, showing the flow of data and knowledge through the system.

The FineSec framework consists of eight interconnected components organized in a pipeline that transforms raw vulnerability data into trained, deployable models capable of accurate vulnerability detection. Figure~\ref{fig:overall} illustrates the complete architecture, showing the flow of data and knowledge through the system.

\textbf{Domain Data \& Knowledge Repository:} Our dataset comprises two components: a training set and a test set. For the training set, we systematically collected vulnerable code snippets and their corresponding fixed versions from open-source repositories, ensuring both authenticity and representativeness. These raw samples underwent rigorous preprocessing and vulnerability labeling, followed by data distillation, to produce domain-optimized training data.
When constructing the test set, we considered two key factors: (1) synthetic test cases designed to target edge conditions, and (2) real-world C/C++-relevant entries in the CWE Top 25 and other high-frequency categories. This hierarchical approach ensures a comprehensive evaluation of both the generalization capability and the real-world detection.

\textbf{Multi-Agent Knowledge Distillation Engine:} The model learns vulnerability patterns through supervised fine-tuning (SFT) with causal language modeling on our distilled dataset. We employ parameter-efficient adaptation using 8-bit quantization and Low-Rank Adaptation (LoRA), freezing the base model while injecting trainable low-rank matrices into the attention layers to efficiently capture security-relevant features. Gradient checkpoint and mixed precision training are applied to maintain computational efficiency, enabling the model to develop precise vulnerability detection capabilities while operating within practical resource constraints. This comprehensive and efficient training strategy enables the model to acquire domain-specific security knowledge, leveraging limited resources to enhance its understanding of vulnerability patterns.

\textbf{Parameter-Efficient Training Module:} The model outputs include quantitative performance metrics and detailed vulnerability detection results, which serve as the basis for assessing overall effectiveness. The experiments are further analyzed across multiple dimensions, with comparisons spanning different model architectures, model parameter scales, real versus synthetic datasets, and diverse vulnerability categories. Through such comprehensive comparisons, we can clarify the model’s capabilities in vulnerability detection across various dimensions, thus identifying its strengths and weaknesses, guiding subsequent optimization directions.

\textbf{Inference \& Evaluation Framework:} This component provides comprehensive evaluation capabilities for trained models, including both offline evaluation on held-out test sets and online evaluation on new code samples. In addition to the ability of the testing model to identify vulnerabilities in the code, the evaluation framework also includes capabilities for analyzing model performance across different vulnerability types, code complexity levels, and programming constructs. This detailed analysis enables identification of model strengths and weaknesses, guiding further training and improvement efforts.

\textbf{Quality Assurance Gate:} This critical component implements a comprehensive validation process that determines whether trained models meet quality standards for deployment. The gate evaluates models across multiple dimensions, including detection accuracy, inference speed, resource consumption, and robustness to adversarial inputs.
The quality gate uses predefined thresholds for each evaluation metric, and models must meet all requirements before being approved for deployment, maintaining the framework's reputation for accuracy and dependability.

\textbf{Model Registry \& Deployment System:} This component manages the lifecycle of trained models, including versioning, documentation, deployment, and monitoring. Successfully validated models are registered with comprehensive metadata, including training data characteristics, performance metrics, and deployment guidelines.
The deployment system supports both batch and online inference modes, enabling integration with various development workflows and security analysis pipelines. It includes monitoring capabilities that track model performance in production environments and alert administrators to potential issues or degradation.

\textbf{Sustainable Learning Mechanisms:} To ensure long-term adaptability and generalization, the model is designed to continuously benefit from newly extracted data and incorporate additional vulnerability types, including those in different programming languages. This is achieved through modular fine-tuning strategies and incremental updates, enabling the model to retain previously acquired knowledge while effectively integrating new information. Such a design supports ongoing improvement and robustness in real-world vulnerability detection scenarios, where threats and code patterns frequently evolve.

\textbf{Shared Knowledge Base:} This central component serves as the repository for all knowledge generated and refined by the framework. It stores training data, model artifacts, evaluation results, and domain expertise in a structured format that enables efficient retrieval and reuse.

\subsubsection{Component Interactions and Data Flow}

The FineSec architecture is designed as a pipeline where each component builds upon the outputs of previous components while contributing to the overall knowledge base. The data flow begins with raw vulnerability data and domain knowledge, which are processed by the multi-agent distillation engine to generate high-quality training examples.

These training examples are then used by the parameter-efficient training module to fine-tune student models, which are subsequently evaluated by the inference and evaluation framework. Models that pass the quality assurance gate are registered and deployed, while their performance data feeds back into the continuous learning engine for ongoing improvement.

This cyclical process ensures that the framework continuously improves its capabilities while maintaining high standards for model quality and reliability. The shared knowledge base serves as the central repository that enables knowledge sharing and reuse across all components, maximizing the efficiency and effectiveness of the entire system.

\subsection{Muti-Agent Knowledge Distillation}
\label{subsec:multi-agent}
The knowledge generation and distillation process forms the core of the FineSec framework. It transforms raw vulnerability data into high-quality training examples that capture both the technical aspects of vulnerabilities and the reasoning processes used by security experts to identify them. This process leverages the capabilities of large teacher models to generate comprehensive, pedagogically effective training data for smaller student models.

Knowledge distillation is a pivotal technique for transferring domain-specific capabilities from large, computationally intensive models to smaller, lightweight architectures, thereby enabling efficient deployment without sacrificing detection accuracy. To enhance the vulnerability detection capacity of smaller LLMs, we construct a distillation-oriented dataset by leveraging a powerful teacher model \emph{GPT-4o} as the source of expert knowledge.

Specifically, we prompt the teacher model by integrating three key strategies: advanced instruction design to guide task orientation, expert knowledge to enhance the model’s understanding of vulnerability background and context, and Chain-of-Thought (CoT) reasoning to enable step-by-step logical deduction in identifying and articulating vulnerabilities. This integrated approach supports the systematic generation of vulnerability-centric code examples. Each instance contains: (1) a well-defined vulnerability label (CWE tag), (2) minimal yet expressive code snippets that encapsulate the vulnerable pattern, and (3) a natural language rationale explaining the root cause of the vulnerability. To ensure clarity and pedagogical effectiveness, the generated examples are simplified and explicitly annotated, highlighting both the vulnerable lines and the corresponding exploit conditions.

\begin{figure}[t]
    \centering
    \includegraphics[width=0.7\linewidth, height=0.8\linewidth, keepaspectratio]{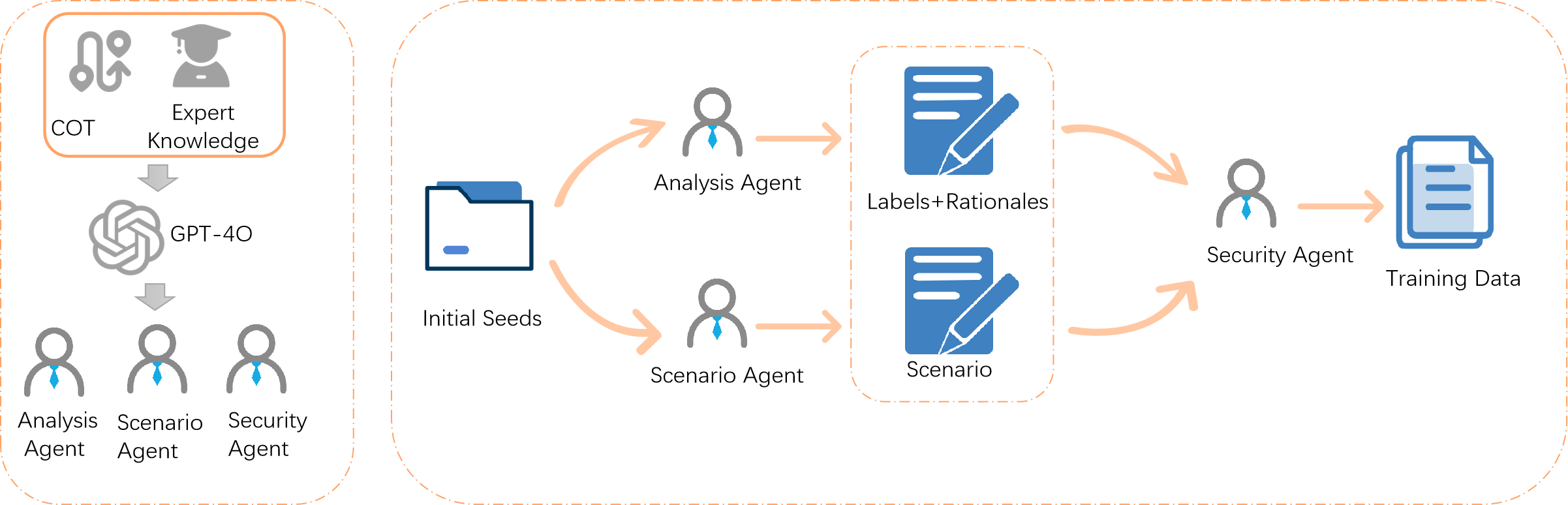}
    \caption{Process of Data Distillation}
    \label{fig:distill}
\end{figure}

To facilitate the creation of high-quality, vulnerability-labeled training data, we employ a multi-agent conversational approach to simulate a role-based virtual dataset-engineering organization: the \textit{Analysis Agent}, the \textit{Scenario Agent}, and the \textit{Security Agent}, each serving distinct yet complementary roles. Detailed process for these agents are presented in Figure~\ref{fig:distill}. Each agent has specific capabilities and roles, specializing in different areas of data generation, including knowledge distillation, vulnerability identification, and code generation. These professional agents conduct in-depth analysis in their respective fields under gradual guidance, resulting in more accurate and comprehensive results. This method enables agents to challenge and supplement each other’s perspectives, leading to a more balanced and thorough analysis.\cite{11121619, 43} 

The roles of these agents are described as follows:

\begin{itemize}
    \item \textbf{Analysis Agent:} The Analysis Agent serves as the primary vulnerability identification component, responsible for analyzing code snippets and generating detailed vulnerability assessments. This agent is equipped with extensive knowledge of vulnerability patterns, CWE taxonomies, and security analysis techniques.
    \item \textbf{Scenario Agent:} The Scenario Agent complements the Analysis Agent by providing contextual information about code usage, intended functionality, and realistic deployment scenarios. This contextual information is crucial for understanding how vulnerabilities might be exploited in real-world environments and for generating training examples that reflect practical security concerns.
    \item \textbf{Security Agent:} The Security Agent synthesizes the outputs of the Analysis and Scenario Agents to generate new code examples that demonstrate specific vulnerability patterns in realistic contexts. This agent serves as the creative component of the multi-agent system, producing diverse training examples that capture the nuances of vulnerability manifestation across different coding styles and application domains.
\end{itemize}

The detailed algorithm is shown as Algorithm \ref{alg:distillation}, where the three agents collaborate to distill vulnerability knowledge, ultimately producing a high-quality labeled dataset for model fine-tuning. The resulting corpus, consisting of paired vulnerability-labeled code and explanatory rationales, is used for parameter-efficient fine-tuning of student models. By internalizing both vulnerability patterns and their underlying reasoning structures, the distilled student models achieve improved detection performance across diverse CWE categories while maintaining computational efficiency.

\begin{algorithm}[ht]
\caption{Multi-Agent Knowledge Distillation for Vulnerability Detection}
\label{alg:distillation} 
\begin{algorithmic}[1]
\REQUIRE Unlabeled code snippets $\mathcal{D} = \{x_i\}_{i=1}^N$
\ENSURE Labeled distilled dataset $\mathcal{D}_{\text{distilled}}$
\STATE \textbf{Initialize} agents with parameters and knowledge bases: 
\STATE \quad Analysis Agent: $\mathcal{A}_{\text{analysis}}(\Theta_A, \mathcal{K}_A)$
\STATE \quad Scenario Agent: $\mathcal{A}_{\text{scenario}}(\Theta_S, \mathcal{K}_S)$  
\STATE \quad Security Agent: $\mathcal{A}_{\text{security}}(\Theta_C, \mathcal{K}_C)$
\STATE \textbf{Initialize} empty dataset $\mathcal{D}_{\text{distilled}} \leftarrow \emptyset$
\FOR{$x_i \in \mathcal{D}$}
    \STATE \COMMENT{Step 1: Generate rationale and label via Analysis Agent}
    \STATE $(r_i, y_i) \leftarrow \mathcal{A}_{\text{analysis}}(x_i; \Theta_A, \mathcal{K}_A)$ \quad $r_i$: vulnerability rationale, $y_i$: CWE label
    
    \STATE \COMMENT{Step 2: Generate scenario description via Scenario Agent}
    \STATE $s_i \leftarrow \mathcal{A}_{\text{scenario}}(x_i; \Theta_S, \mathcal{K}_S)$ \quad $s_i$: natural language context/scenario
    
    \STATE \COMMENT{Step 3: Synthesize vulnerable code via Security Agent}
    \STATE $\tilde{x}_i \leftarrow \mathcal{A}_{\text{security}}(r_i, y_i, s_i; \Theta_C, \mathcal{K}_C)$ \quad $\tilde{x}_i$: synthesized vulnerable code snippet
    
    \STATE \COMMENT{Step 4: Validate syntactic validity and minimality}
    \IF{$\Psi(\tilde{x}_i) = 1$}
        \STATE \quad $\Psi$ validates compilability and minimality
        \STATE \quad \textbf{Add} $(\tilde{x}_i, y_i)$ to $\mathcal{D}_{\text{distilled}}$
    \ENDIF
\ENDFOR
\STATE \RETURN $\mathcal{D}_{\text{distilled}}$
\end{algorithmic}
\end{algorithm}
    
\subsection{Three-Stage Training Pipeline}
\label{subsec:three_stage_training}
Our framework transforms LLMs into domain-specialized models for vulnerability detection through three key stages. In the initial pre-pretraining phase, the base model’s cognitive ability for security semantics is carefully optimized, laying a solid foundation for domain adaptation. Next, fine-tuning is performed using the data distillation output as the training dataset, enabling the model to learn vulnerability-specific features. Finally, alignment is achieved through instruction tuning, connecting domain knowledge with task execution to enable precise vulnerability identification and analysis. This phased approach systematically cultivates domain models capable of conducting expert-level security analysis.

\begin{figure}[t]
    \centering
    \includegraphics[width=0.6\linewidth]{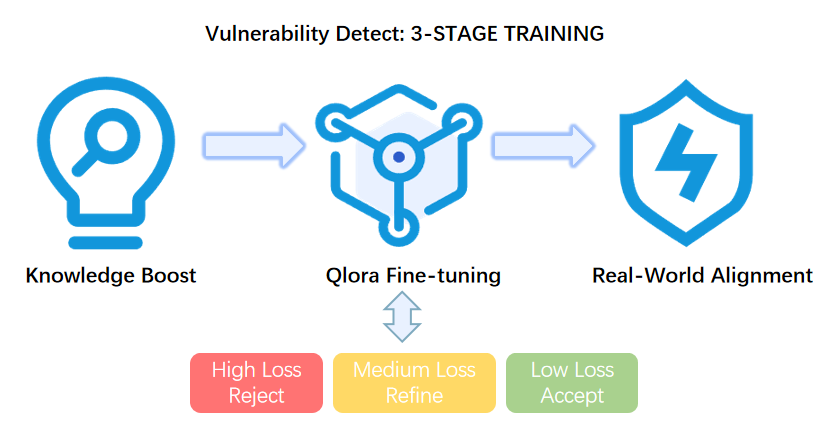}
    \caption{Three-Stage Training Pipeline of FineSec}
    \label{fig:overall}
\end{figure}

\paragraph{Stage 1: Foundational Pre-pretraining}
The first stage prepares the base model by enhancing its understanding of security-specific language. This is done by expanding its vocabulary with key security terms like \texttt{integeroverflow} and \texttt{accesscontrol}. Treating these terms as single units prevents them from being broken down, which improves the model's ability to efficiently encode and comprehend security concepts.

\paragraph{Stage 2: Iterative Fine-tuning with Quality Control}
This stage develops the model's vulnerability detection skills through a unique, iterative process as Algorithm~\ref{alg:iterative_enhancement}. 

Initial Training: The model is fine-tuned on a high-quality dataset containing code snippets, vulnerability labels, and expert explanations. This stage leverages the high-quality training data $D^{(0)} = \{x_i, y_i, r_i\}_{i=1}^N$ generated by the multi-agent knowledge distillation process, where $x_i$ represents code snippets, $y_i$ denotes vulnerability labels, and $r_i$ contains rational explanations.

Iterative Refinement: The model's performance is evaluated based on a loss score. This score triggers one of three actions based on two predefined thresholds: $L_h$ (upper threshold, exceeding this indicates ineffective learning) and $L_l$ (lower threshold, falling below this signifies satisfactory performance): 
\begin{itemize}
    \item {High-loss ($L_j$ >$L_h$)}: Models with $L_j > L_h$ are discarded as it's not learning effectively.
    \item {Medium-loss ($L_l \leq L_j \leq L_h$)}: The model is kept, but human experts refine the training data to improve it for the next training iteration.
    \item {Low-loss ($L_j < L_l$)}: Models achieving $L_j < L_l$ meet the quality standards, terminating the iterative process with satisfactory modelscomplies with the quality standard and the process is completed.
    \item {Efficiency}: This stage employs QLoRA (Quantized Low-Rank Adaptation), a parameter-efficient fine-tuning method that reduces computational complexity via weight quantization and low-rank matrix decomposition. This preserves model capacity while minimizing memory footprint and training latency, enabling scalable iterative refinement.
\end{itemize}

\begin{algorithm}[ht]
\caption{Iterative Enhancement Process}
\label{alg:iterative_enhancement}
\begin{algorithmic}[1]
\STATE \textbf{Input:}
\STATE \quad Initial fine-tuning dataset $D^{(0)} = \{x_i, y_i, r_i\}_{i=1}^N$
\STATE \quad Domain-adapted Models $M_s^{(0)} = \{m_j\}_{j=1}^M$
\STATE \quad Label prediction loss threshold $L_b$
\STATE \quad Dual-thresholds for function loss $L_l$ and $L_h$ ($L_l < L_h$)
\STATE \quad Maximum iterations $K_{max}$
\STATE
\STATE \textbf{Output:}
\STATE \quad Satisfactory fine-tuned model set $M_{ft}^{(K)}$
\STATE \quad Satisfactory dataset $D_{train} = D^{(K)}$
\STATE
\STATE Initialize $k = 0$, $satisfactory\_found = \text{false}$
\STATE $M_{ft}^{(K)} \leftarrow \emptyset$
\REPEAT
    \STATE Fine-tune models in $M_s^{(k)}$ using $D^{(k)}$ to get $M_{ft}^{(k)}$
    \STATE Initialize $M_s^{(k+1)} \leftarrow \emptyset$, $dataset\_modified \leftarrow \text{false}$
    \STATE
    \FOR{each model $m_j$ in $M_{ft}^{(k)}$}
        \STATE Compute combined losses $L$ for all inputs in $D^{(k)}$:
        \STATE Initialize $L_j \leftarrow 0$
        \FOR{each input $x_i$ in $D^{(k)}$}
            \STATE Model $m_j$ outputs $\hat{y}_i$ and $\hat{r}_i$ using $x_i$
            \STATE Update $L_j$ using $\hat{y}_i$, $\hat{r}_i$, $y_i$, $r_i$
        \ENDFOR
        \STATE $L_j \leftarrow L_j / N$ \COMMENT{Normalize by dataset size}
        \STATE
        \IF{$L_j \leq L_l$}
            \STATE Add $m_j$ to $M_{ft}^{(K)}$
            \STATE $satisfactory\_found \leftarrow \text{true}$
        \ELSIF{$L_j \leq L_h$}
            \STATE Add $m_j$ to $M_s^{(k+1)}$
            \IF{not $dataset\_modified$}
                \STATE Manually modify $D^{(k)}$ to form $D^{(k+1)}$
                \STATE $dataset\_modified \leftarrow \text{true}$
            \ENDIF
        \ELSE
            \STATE \COMMENT{Model $L_j > L_h$ is discarded}
        \ENDIF
    \ENDFOR
    \STATE
    \IF{not $dataset\_modified$}
        \STATE $D^{(k+1)} \leftarrow D^{(k)}$ \COMMENT{Preserve dataset if no modifications}
    \ENDIF
    \STATE
    \STATE Increment $k$
\UNTIL{$satisfactory\_found$ or $k \geq K_{max}$ or $M_s^{(k)} = \emptyset$}
\STATE
\STATE $D_{train} \leftarrow D^{(k)}$
\end{algorithmic}
\end{algorithm}

This combined approach of iterative quality-aware enhancement and parameter-efficient fine-tuning enables effective adaptation of LLMs for vulnerability detection while maintaining computational feasibility and ensuring high-quality model outputs.

\paragraph{Stage 3: Practical Alignment}
The final stage ensures the model's output is practical for real-world use. It focuses on aligning the model's responses so they are accurate, useful, and correctly formatted. This step bridges the gap between the model's technical knowledge and the specific needs of security analysis workflows, making it ready for deployment.

\begin{figure}[h]
    \centering
    \includegraphics[width=0.65\linewidth, height=0.65\linewidth, keepaspectratio]{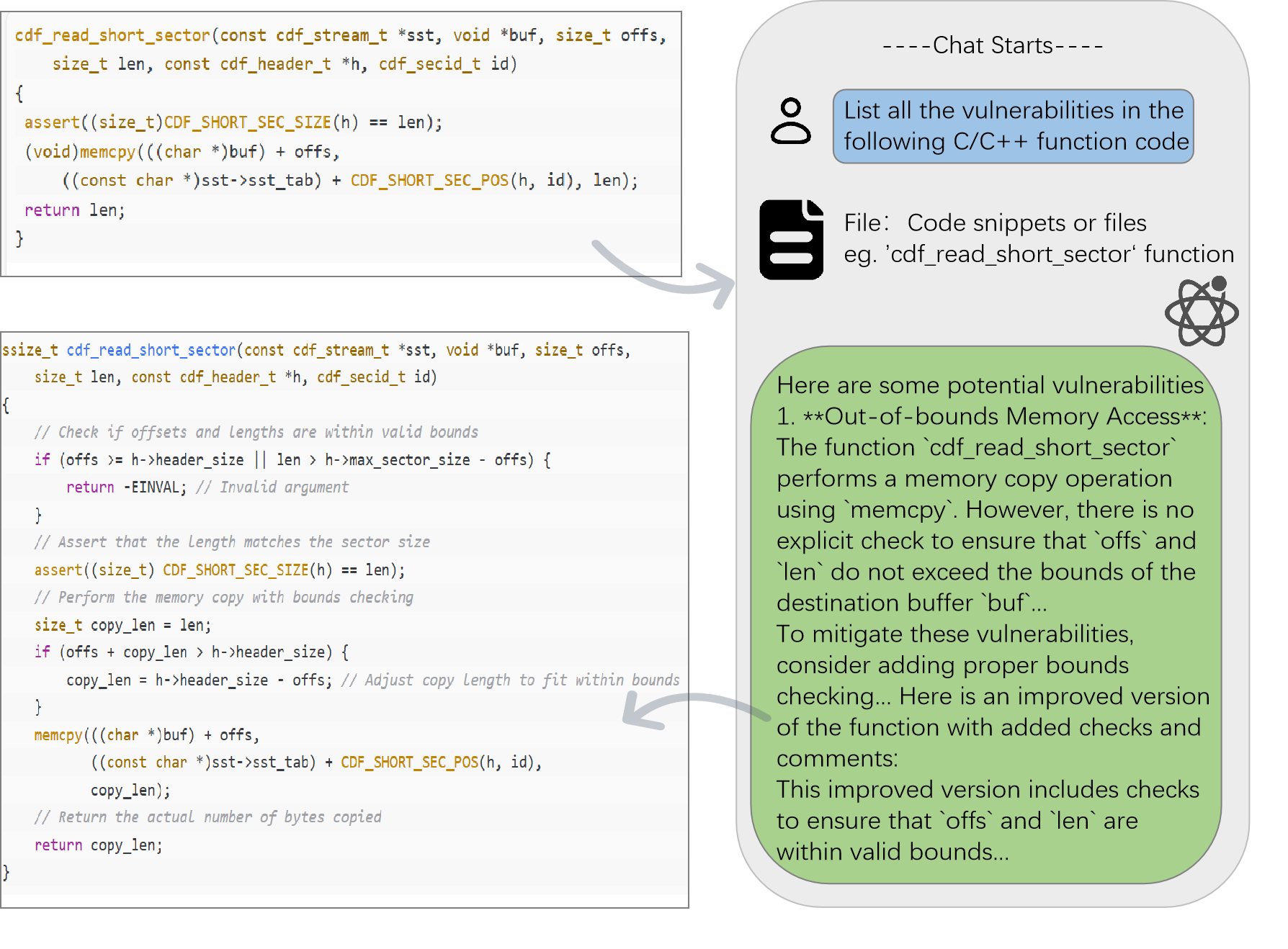}
    \caption{A Real Example of Model Interaction}
    \label{fig:interaction}
\end{figure}

As shown in Figure~\ref{fig:interaction}, the vulnerability analysis report generated by the Llama-3.2-3B model systematically identifies security vulnerabilities, including their types, locations, root causes. In addition to detection, the report also produces patch code to fix the vulnerabilities and offers actionable repair recommendations to support risk mitigation and prevention.

\subsection{Illustrative Walkthrough (C/C++): CWE-190 Uncontrolled Format String}

\paragraph{Stage 1: Data \& Knowledge}
We curate canonical integer overflow cases and annotate them with CWE-190. The knowledge base records secure patterns (e.g., range checks before arithmetic operations) and the associated rationale (preventing integer values from exceeding their maximum representable limits).

\paragraph{Stage 2: Multi-Agent Distillation}
From heterogeneous sources, the multi-agent distiller produces compact training items that isolate the core fault.

\begin{lstlisting}[language=C, caption={Distilled training item for CWE-190 (unsafe)}, label={lst:cwe190-train-unsafe}]
#include <stdio.h>
void process_size(int user_size) {
    int buffer_size = 1000;
    int total = buffer_size + user_size;  // vulnerable: no overflow check
    char buffer[total];                   // risk of integer overflow
}
\end{lstlisting}

A paired secure pattern is stored as a counterexample:

\begin{lstlisting}[language=C, caption={Secure pattern (overflow check)}, label={lst:cwe190-train-safe}]
#include <stdio.h>
#include <limits.h>
void process_size_safe(int user_size) {
    int buffer_size = 1000;
    // Fixed: check for potential overflow before calculation
    if (user_size > INT_MAX - buffer_size) {
        printf("Error: Integer overflow risk\n");
        return;
    }
    int total = buffer_size + user_size;
    char buffer[total];
}
\end{lstlisting}

\paragraph{Stage 3: Training (PEFT)}
We fine-tune with PEFT (e.g., QLoRA) using instruction-style exemplars. The output format is constrained to a structured schema \texttt{\{issue, taxonomy, severity, locations, patch\}}.

\paragraph{Stage 4: Inference \& Evaluation}
On a held-out C/C++ validation set, the model achieves class-level performance consistent with accuracy, cost, time and so on.

\paragraph{Stage 5: Model Selection}
Only when all thresholds such as detection accuracy, inference speed, and resource consumption are met, will the candidate be frozen, recorded with a model card, and then proceed with subsequent testing and experiments.

\paragraph{Stage 6: Target Code Test (post-release)}
The released model is applied to the user’s target C/C++ snippet:

\begin{lstlisting}[language=C, caption={Target snippet analyzed post-release}, label={lst:cwe190-target}]
void copy_data(int count) {
    int buffer_size = count * 4;  
    char *data_buffer = malloc(buffer_size);
    
    if (data_buffer) {
        printf("Allocated %d bytes\n", buffer_size);
        free(data_buffer);
    }
}
\end{lstlisting}

The model produces a structured finding:
\begin{lstlisting}[language={} , caption={Model output (structured, abridged)}, label={lst:cwe190-output}]
{
  "target": "target.c:copy_data",
  "detections": [{
    "issue": "Integer overflow in arithmetic operation",
    "taxonomy": {"CWE": "CWE-190"},
    "locations": [{"file": "target.c", "lines": "2"}],
    "rationales": [
      "The multiplication 'count * 4' may exceed the maximum value representable by int (INT_MAX).",
      "Large user-controlled 'count' values can trigger overflow, leading to incorrect buffer_size calculation.",
      "Subsequent malloc() with an overflowed size may allocate insufficient memory, causing buffer underflow."
    ],
    "patch": {
      "strategy": "Add overflow check before calculation",
      "diff": [
        "+        if (count > INT_MAX / 4) {",
        "+            fprintf(stderr, \"Overflow risk\\n\");",
        "+            return;",
        "+        }",
        "         int buffer_size = count * 4;"
      ]
    }
  }]
}
\end{lstlisting}
\paragraph{Outputs \& Continuous Learning}
The report and patch suggestion are delivered to the user. Confirmed before/after pairs (Listings~\ref{lst:cwe190-train-unsafe}--\ref{lst:cwe190-target}) and any deployment feedback are fed back to the knowledge base, and periodically incorporated into incremental PEFT updates.

\section{Evaluation}
To assess the effectiveness of our fine-tuned LLMs in vulnerability detection, we apply FineSec to enhance six models, include GPT-4o and DeepSeek-R1 for comparison, and evaluate them on a test dataset covering 30 C/C++ CWE types. In this section, we state the research questions, describe the experimental setup, including dataset statistics, model configurations, and the training environment, and then present a detailed performance analysis. We then discuss the results in the context of the research questions posed earlier in the paper.

\subsection{Research Questions}
Our evaluation aims to address the following research questions:

\begin{itemize}

\item \textbf{RQ1: What are the key factors shaping the security vulnerability detection capabilities of LLMs?}
We conducted a controlled comparative analysis: we evaluated multiple large language models varying in parameter sizes and architectural designs across a diverse set of vulnerability detection datasets encompassing both real-world vulnerability corpora and synthetically generated samples. This allowed us to analyze the factors influencing models' vulnerability detection capabilities from multiple perspectives, including dataset properties (real-world versus synthetic), model parameter scales, and architectural designs.

\item \textbf{RQ2: How does FineSec affect LLM performance?}
We propose FineSec, a framework designed to train LLMs in specific domains while minimizing computational overhead, thus adapting them more effectively to vulnerability detection tasks. We apply FineSec to five representative LLMs and the empirical results show that it significantly improves the detection performance on real-world vulnerability datasets. These findings highlight FineSec's effectiveness in improving domain adaptability and security awareness of general-purpose LLMs.

\item \textbf{RQ3: How does the performance of LLMs vary across different CWE categories?} We conducted a systematic analysis of the experimental results from the perspective of vulnerability taxonomy. The evaluation covers more than 30 vulnerability types, categorized into five major classes including Memory Safety, Input Validation \& Injection, System Resource \& Logic Errors, Permissions \& Access Control, and Cryptography \& Information Leakage, which allows us to uncover fine-grained patterns and strengthens the comprehensiveness and depth of our vulnerability evaluation.

\item \textbf{RQ4: Does FineSec have the ability to detect new vulnerabilities?}
The ability to discover previously unknown vulnerabilities is a critical metric to evaluate a model’s detection capability. Through systematic analysis of model-generated vulnerability reports, we observed that FineSec demonstrates high sensitivity to novel vulnerability patterns. In particular, it successfully identified nine previously unreported vulnerabilities in C/C++ language code, underscoring its potential to advance proactive software security.

\end{itemize}

\subsection{Experimental Setup}
\subsubsection{Dataset Selection} 

In this study, we focus on the C/C++ programming language due to its widespread use in the development of server-side, Android, and web applications—domains that are particularly susceptible to security vulnerabilities. Moreover, C/C++ has a long development history and a substantial number of publicly disclosed CVEs across various projects, providing rich resources for vulnerability analysis.

We require a dataset that satisfies several key criteria:

\begin{itemize}
\item Each benchmark should include relevant vulnerability metadata, such as CWE ID, CVE ID, and the vulnerable function.

\item Detailed vulnerability descriptions should be available to assist the model in learning and subsequent evaluation.

\item The project length should be suitable for model training: code that is too short may lack sufficient context, while overly long code may exceed input limits or introduce noise, reducing training efficiency.

\item Functions containing vulnerabilities should have corresponding patched code, enabling direct comparison to better identify the root cause of the vulnerability.
\end{itemize}
\begin{table}[h]
  \caption{The overview of the test datasets} 
  \label{tab:datasets}
  \begin{tabular}{cccccc}
    \toprule 
    Type & Dataset & Obtaining Method & Size & Vul/Non-Vul & CWEs \\ 
    \midrule  
    Synthetic & SARD Juliet (C/C++) & Synthetically Generated & 81,280 & 40,640/40,640 & 118 \\
    Real World & CVEFixes & CVE Commit Diff & 61,638 & 17,725/43,913 & 180 \\ 
    \bottomrule 
  \end{tabular}
\end{table}

Based on these requirements, we select two complementary vulnerability datasets: one derived from a synthetic dataset and the other collected from real-world software projects, as shown in  Table~\ref{tab:datasets}. Due to the small number of samples for some vulnerabilities in the dataset, to ensure the credibility of the test results, we selected 30 vulnerabilities with larger sample sizes for the experiment. These include the top 25 C/C++-related vulnerabilities and the most frequently occurring ones, focusing on detection from two   perspectives: harmfulness and high frequency.

\begin{itemize}  

\item \textbf{Juliet C/C++ Test Suite 1.3:}
This dataset \cite{25}, developed by the National Security Agency’s Center for Assured Software, is synthetically constructed using automated tools to generate C/C++ code fragments that systematically embed common vulnerability patterns.

\item \textbf{CVEfixes Dataset:}
In contrast, the CVEfixes dataset \cite{27} is compiled from real-world vulnerabilities and includes actual code fixes implemented to address identified security issues in various open-source projects and enterprise software systems. This dataset reflects genuine security challenges encountered in production environments, providing practical examples of vulnerabilities and their corresponding resolutions.

\end{itemize}

\subsubsection{Pre-processing of test data} 

Pre-processing test data is a key step in ensuring the effectiveness of vulnerability detection task evaluation. Its core goal is to improve data quality, eliminate interference factors, and ensure data compatibility with task requirements through a series of standardized operations. We perform the following data preprocessing steps on the above datasets:

\begin{itemize}

\item \textbf{Length filtering:} To optimize training efficiency and resource utilization, we exclude code files larger than 32,765 bytes and remove functions shorter than three lines due to insufficient vulnerability detection information. Ultra-long code often contains a large amount of irrelevant content (e.g., redundant logic, duplicate code, comments), which models may mistakenly treat as important patterns. This can lead to learning irrelevant statistical correlations rather than true vulnerability features, while also increasing resource consumption and reducing efficiency. Conversely, overly short code typically lacks sufficient vulnerability-related information, making it difficult to capture and analyze meaningful patterns.

\item \textbf{Duplicate data deletion:} Studies have shown that duplicated examples can distort model outputs and reduce generation diversity\cite{46}. We remove duplicate samples from the dataset to prevent them from negatively affecting model learning, thereby improving both training efficiency and the generalizability of large models.

\item \textbf{Information hiding:} We remove comments and code containing CWE-related information to prevent straightforward identification of vulnerabilities. Otherwise, the model may simply memorize keywords in tags or comments (e.g., \texttt{/* buffer */}) instead of learning the actual vulnerability patterns present in the code.

\item \textbf{Keyword replacement:} We replace terms such as \texttt{“BAD”}, \texttt{“GOOD”}, \texttt{“VULN”}, and \texttt{“PATCHED”} in function names with the placeholder \texttt{“func”} to prevent the model from inferring vulnerabilities directly from such keywords.

\end{itemize}

\subsubsection{Selected Models} 

A wide range of LLMs are available, and our selection for this study was guided by the following criteria:
\begin{itemize}
\item \textbf{Adaptability to code tasks:} The model must demonstrate strong code understanding capabilities and show potential for improved performance on vulnerability detection tasks after fine-tuning.

\item \textbf{Model size:} Since parameter count directly correlates with computational cost during training, memory usage during deployment, and inference speed in real-world workflows, we prioritize models with around 10B parameters. This focus allows us to highlight the efficiency of smaller models in practical security scenarios, where resource constraints (e.g., limited GPU memory in edge environments or cost-sensitive deployment) often restrict the use of very large models.

\item \textbf{Maintainability and extensibility:} The model should support seamless updates, retraining, and incremental learning to quickly adapt to emerging vulnerabilities and evolving attack methods. This includes compatibility with modular fine-tuning techniques (such as LoRA) that enable efficient adaptation without full retraining, as well as architectural flexibility to incorporate new data modalities (e.g., static analysis graphs or runtime logs) as they become relevant to vulnerability detection.

\item \textbf{Open-source availability and community support:} Preference is given to open-source models with active community engagement, such as the Mistral and LLaMA series, to leverage shared resources and development support. Additionally, active community support—manifested through frequent updates, shared fine-tuning recipes, and troubleshooting resources—facilitates collaborative development, accelerates debugging, and ensures long-term sustainability of the model for evolving security use cases.
\end{itemize}

\begin{table}[h]
  \caption{Overview of the Selected Language Models}
  \label{tab:llms}
  \centering
  \begin{tabular}{llr}
    \toprule
    \textbf{Model Series} & \textbf{Model Variant} & \textbf{Parameters} \\
    \midrule
    LLaMA & 3.1-8B-Instruct & 8B \\
    LLaMA & 3.2-3B-Instruct & 3B \\
    LLaMA & 3.2-1B-Instruct & 1B \\
    Qwen & 2.5-7B-Instruct & 7B \\
    Mistral & 7B-Instruct & 7B \\
    DeepSeek & R1 & 236B (MoE) \\
    GPT & 4o & 720B \\
    \bottomrule
  \end{tabular}
\end{table}

Following these criteria, we selected Llama-3.1-8B-Instruct, Llama-3.2-3B-Instruct, and Llama-3.2-1B-Instruct from the LLaMA series, Qwen2.5-7B-Instruct, Mistral-7B-Instruct, GPT-4o, and DeepSeek-R1 as our reference models (Table~\ref{tab:llms}).

\subsection{Evaluation Settings} 
\subsubsection{Evaluation Criteria} 
To assess the effectiveness of each tool, we employ standard classification metrics. In this study, detection results are framed as a binary classification problem, with outcomes categorized as follows:

\begin{itemize}
\item \textbf{True Positive (TP):} The tool correctly detects a vulnerability in a contract where one exists.
\item \textbf{False Positive (FP):} The tool incorrectly reports a vulnerability in a contract where none exists.
\item \textbf{False Negative (FN):} The tool fails to detect an existing vulnerability in a contract.
\item \textbf{True Negative (TN):} The tool correctly determines that a contract contains no vulnerability.
\end{itemize}

Accuracy measures the proportion of correct predictions—whether identifying a vulnerability or confirming its absence—and is calculated as:

\begin{equation}
\text{Accuracy} = \frac{\text{TP} + \text{TN}}{\text{TP+TN+FP+FN}}
\end{equation}

\subsubsection{Implementation Details} 
\begin{figure}[h]
    \centering
    % 左侧训练损失图：固定高度5cm，宽度自适应子图容器
    \begin{minipage}[t]{0.45\linewidth}
        \centering
        \includegraphics[width=\linewidth, height=3cm]{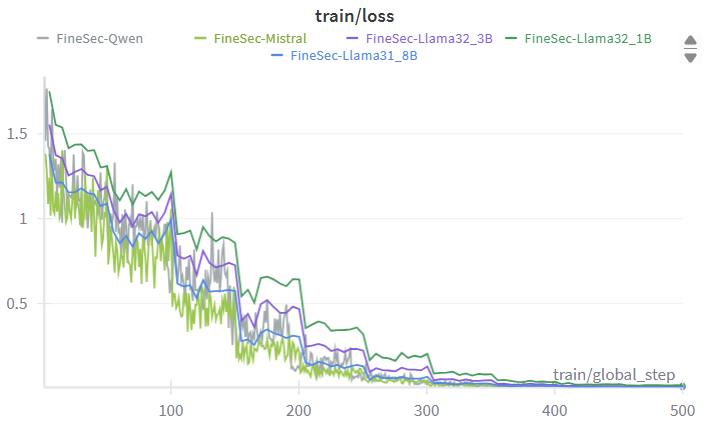}  % 新增height=5cm控制高度
        \caption{Training Loss of process}
        \label{fig:image1}
    \end{minipage}%
    \hspace{0.05\linewidth}  % 两图间距（保持不变）
    % 右侧梯度范数图：与左侧高度一致，确保视觉对称
    \begin{minipage}[t]{0.45\linewidth}
        \centering
        \includegraphics[width=\linewidth, height=3cm]{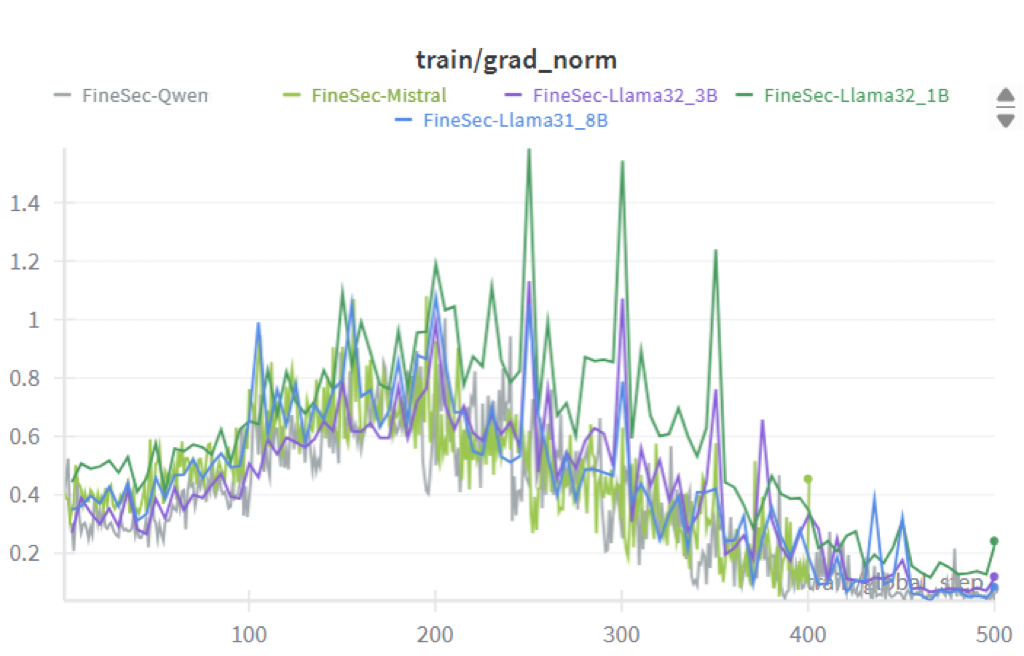}  % 同左侧高度
        \caption{Grad\_norm of process}
        \label{fig:image2}
    \end{minipage}
\end{figure}

We conducted model training on NVIDIA T4 GPUs, each equipped with 16 GB of VRAM. To adapt the model to the training dataset, we dynamically adjusted fine-tuning hyperparameters based on training loss and gradient norms. After multiple rounds of empirical testing, we set the learning rate to 5e-4 and and the total number of training steps to 500, ensuring stable convergence under limited computational resources. To balance efficiency and performance, we used a small batch size, the 8-bit AdamW optimizer, and mixed-precision training (FP16). A linear learning rate schedule with warm-up was applied, and model checkpoints were saved every 50 steps. All training runs were tracked using Weights \& Biases to ensure reproducibility and effective monitoring, as shown in Figures~\ref{fig:image1} and~\ref{fig:image2}.

\subsection{Experimental Results}
\subsubsection{RQ1: Impact of code source on detection performance}

In this section, we first evaluate the ability of GPT-4o to detect vulnerabilities across different code styles in both synthetic and real-world datasets. This analysis examines how variations in coding conventions and patterns affect the model’s effectiveness in identifying vulnerabilities. We then conduct an extensive zero-shot evaluation on eight models to assess their inherent vulnerability detection capabilities. Using zero-shot prompts, we instruct each model to enumerate all vulnerabilities within specified C/C++ code snippets. All models are trained to convergence under identical hardware conditions (NVIDIA T4 GPUs), ensuring that performance differences are not influenced by hardware or training procedure variations. 

Synthetic vs. Real-world Detection: We conducted tests on two distinct datasets to assess the performance of GPT-4o: the Juliet Test Suite for C/C++ 1.3, a synthetic dataset, and the CVEfixes dataset, which is derived from real-world scenarios. By evaluating across these two datasets to comprehensively measure the ability of LLMs to detect vulnerabilities under different conditions. The final test results are presented in Table~\ref{tab:overview}.

\begin{table}[ht]
  \caption{GPT-4o Performance on Different Vulnerability Datasets}
  \label{tab:overview}
  \centering
  \begin{tabular}{lllc}
    \toprule
    \textbf{Data Type} & \textbf{Dataset} & \textbf{Language} & \textbf{Accuracy} \\
    \midrule
    Synthetic & SARD Juliet Test Suite & C/C++ & 90.72\% \\
    Real-world & CVE Fixes Repository & Mixed & 53.33\% \\
    \bottomrule
  \end{tabular}
\end{table}

We found that GPT-4o achieves a high accurcay on synthetic datasets, likely because such datasets are typically generated from known vulnerability patterns and features, have relatively standardized structures and syntax, and are therefore easier for models to recognize. In contrast, real-world datasets exhibit greater code complexity, including style variations and code noise, which increases detection difficulty. Real code often contains complex contextual relationships, whereas the context in synthetic data is comparatively simple.

\begin{figure}[h]
  \centering
  % 新增 height=6cm 控制高度，width 保持原0.9\textwidth
  \includegraphics[width=0.9\textwidth, height=5cm]{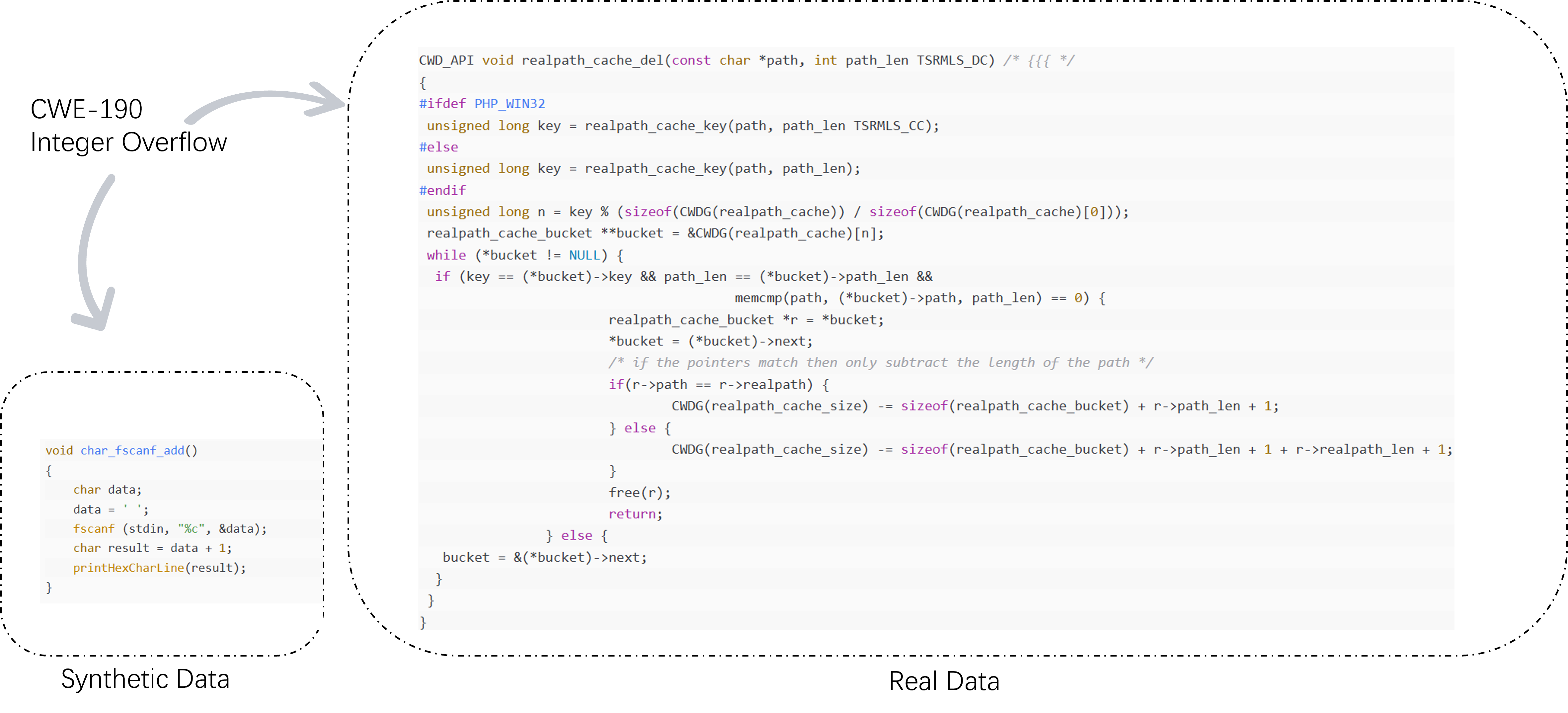} 
  \caption{The difference in style between synthetic code and real code}
  \label{fig:differ}
\end{figure}

For example, Figure~\ref{fig:differ} illustrates stylistic differences in the same vulnerability as it appears in synthetic and real-world datasets. Based on these observations, our subsequent experiments will focus on real-world datasets and aim to improve the ability of various models to detect vulnerabilities in complex, real-world scenarios. 

\begin{figure}[t]
  \centering
  % 新增 height=6cm 控制高度，width 保持原0.9\textwidth
  \includegraphics[width=0.6\textwidth]{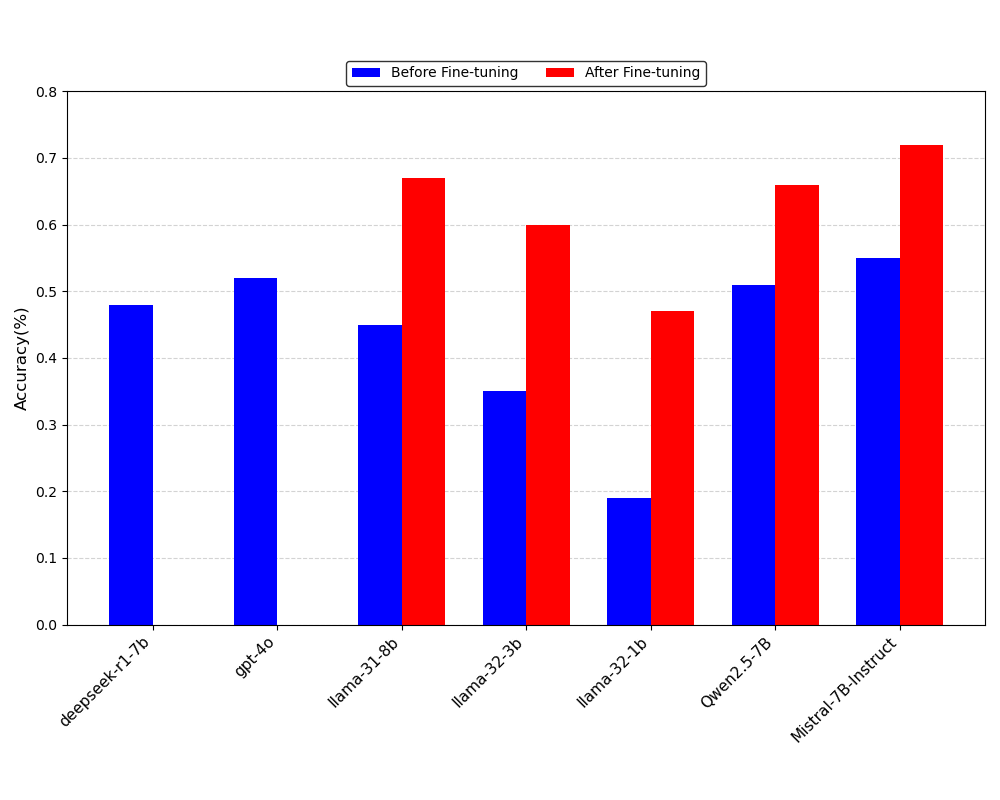} 
  \caption{Models Performance: Before vs. After FineSec}
  \label{fig:perform}
\end{figure}

To address the issue that current models are incompatible with the "complexity noise" in real-world code, we have implemented scenario-specific feature enhancement. Specifically, we extract "scenario-specific features" of real-world vulnerabilities (such as memory operation features in embedded code and input validation features in Web backends) through multi-agent data distillation. These features are then injected into the model via fine-tuning, enabling the model to recognize "real-world vulnerability patterns that are absent in synthetic data". Ultimately, the model achieves a significant improvement in detection rate on synthetic datasets as shown in Figure~\ref{fig:perform}. and enhances the adaptability of baseline models to complex real-world code.

Based on the preliminary detection results, we can see that other two key factors shaping the security vulnerability detection capabilities of LLMs:

Modest Vulnerability Detection Performance Across LLMs: Before FineSec, the vulnerability accuracy of most models is around 50\%. Mistral-7B-Instruct demonstrates a strong ability to handle code-specific challenges. GPT-4o ranks second in vulnerability detection, largely due to its extensive parameter set, which enables comprehensive feature mapping and a better understanding of complex data relationships. After FineSec, the vulnerability accuracy of most models is around 60\%, and llama series and qwen models improve their ability to detect vulnerabilities that surpass the chosen big models.

Impact of Model Size: Despite having fewer parameters compared to GPT-4o, LLaMA-3.1-8B performs remarkably well in vulnerability detection tasks, highlighting its efficiency in code processing. Analysis of the LLaMA model’s parameter scaling indicates that as the number of parameters decreases, the model’s overall ability to detect vulnerabilities diminishes. Higher parameter counts enable the capture of more complex features and relationships, thereby enhancing performance on specialized tasks. Conversely, smaller models, while more efficient and less resource-intensive, may struggle to detect subtle vulnerability patterns in complex scenarios, leading to reduced detection capabilities. Notably, under equivalent conditions for detecting the same vulnerabilities, the 8B model takes three times longer than the 3B model and six times longer than the 1B model, underscoring the need to balance time efficiency and accuracy in future evaluations.

\begin{tcolorbox}[colback=gray!20, colframe=gray!80, sharp corners, boxrule=0.5pt]
\textbf{Answer to RQ1:} Detection performance is influenced by three key factors: \textbf{Code style} – Models achieve high accuracy on structured, synthetic datasets but struggle with the complexity and variability of real-world code. \textbf{Model architecture} – Mistral delivers the best results, followed by GPT-4o and Qwen base models. Llama series show the most significant performance improvement after FineSec. \textbf{Parameter count} – In the LLaMA series, fewer parameters improve processing efficiency but reduce detection accuracy, underscoring the need to balance resource usage, precision, and runtime in vulnerability detection tasks.
\end{tcolorbox}

\subsubsection{RQ2: How does FineSec affect LLM performance?} FineSec enhances LLM performance in vulnerability detection through three consecutive optimization stages: pre-training to establish a security-related knowledge foundation, fine-tuning using distilled training data, and instruction-based alignment to link domain knowledge with detection tasks. We used the same test dataset as the baseline to quantitatively evaluate this optimized model and determine whether its performance had improved. From the results, the model has enhanced its vulnerability detection capabilities in multiple dimensions: it achieves a higher detection rate for vulnerabilities, delves deeper into the root causes and contextual dependencies of detected vulnerabilities, and generates more standardized and actionable vulnerability reports with clear classification, detailed evidence chains, and consistent formatting.

The experimental results in Figure~\ref{fig:perform} demonstrate a significant improvement in the effectiveness of FineSec in accurately identifying vulnerabilities in LLM. Llama-32-1B achieved the largest performance. However, because of its smaller parameter size, its final capability can only be considered moderate. In summary, after applying FineSec, all other models outperformed DeepSeek-R1 and GPT-4o. LLaMA models proved to be particularly well suited for enhancement, with each achieving improvements of more than 20\%, followed by the Mistral models and then the Qwen models. In particular, Mistral demonstrated an excellent vulnerability detection capability, achieving an overall accuracy of 73.3\% and surpassing 80\% accuracy in 11 CWEs across the five major vulnerability categories.

Despite overall improvements, FineSec rarely lowers detection rates for specific vulnerabilities that initially had high accuracy. For example, Mistral model in CWE-476 decreased from 0.98 to 0.84, highlighting a potential trade-off introduced by FineSec. This suggests that enhancing general detection capability may come at the cost of reduced performance in detecting specific, well-identified vulnerabilities.

Through a comparative analysis of vulnerability reports using a representative example, before and after FineSec implementation, we found that both models identified issues at same lines: 10–14. The Mistral base model detected a null pointer exception: `If \texttt{re\_yyalloc} or \texttt{re\_yyrealloc} fails to allocate memory, the function will attempt to dereference a null pointer when calling \texttt{YY\_FATAL\_ERROR}. This could lead to a null pointer dereference.' However, the model processed by FineSec stated: ‘If the allocation of \texttt{yyg->yy\_buffer\_stack} fails, the function calls \texttt{YY\_FATAL\_ERROR} and does not free the previously allocated memory. This could lead to a resource leak if the error is not handled properly.' While the base model detected the immediate crash risk (CWE-476), FineSec additionally identified the resource management flaw (CWE-401) that could lead to progressive system degradation, demonstrating deeper analysis by exposing architectural anti-patterns rather than just symptoms.

Similarly, when comparing other CWE models (which showed reduced performance) before and after FineSec processing, we found that the FineSec-processed model tended to focus more on the stability of the system's long-term behavior as a chain risk, while overlooking certain single-operation or point errors.

\begin{figure}[t]
  \centering
  % 新增 height 限制，同时用 keepaspectratio 保持宽高比，避免变形
  \includegraphics[
    width=0.9\textwidth,  % 保留原宽度（占页面90%）
    height=0.2\textheight,  % 设定最大高度（占页面文本高度40%，可按需调整）
    keepaspectratio        % 关键：保持宽高比，防止拉伸/压缩变形
  ]{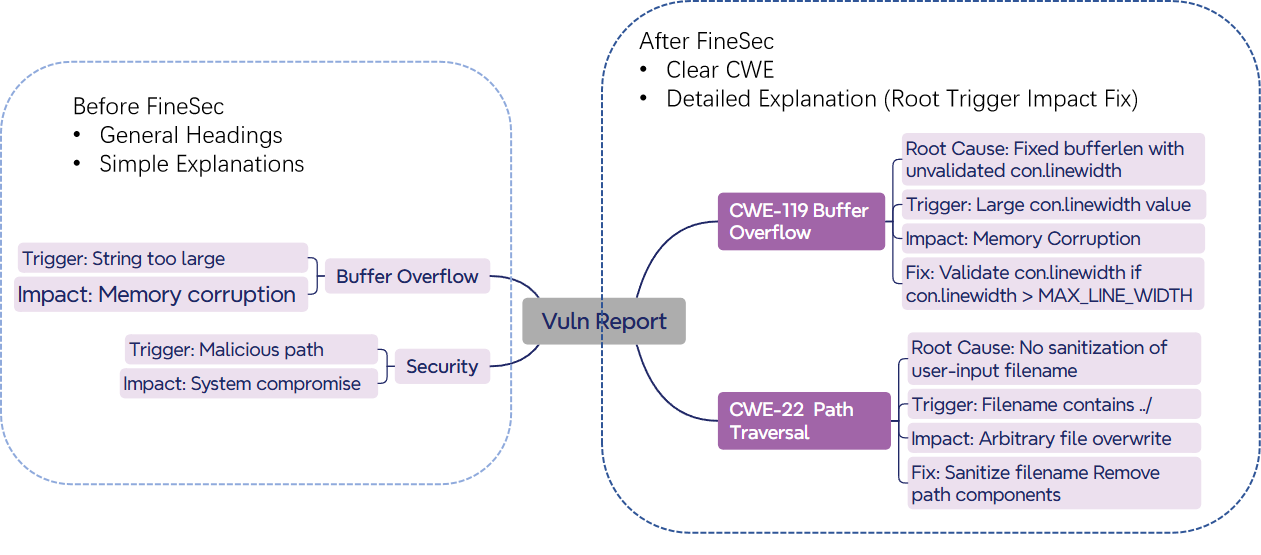} 
  \caption{Vulnerability Report Comparison Before and After FineSec}
  \label{fig:report}
\end{figure}

To address the ambiguity of generic vulnerability reports—where vague descriptions, fragmented evidence, and inconsistent formatting often force security auditors to spend extra effort verifying and organizing information—FineSec-optimized models generate standardized and actionable vulnerability reports by integrating three core design principles: structured classification, traceable evidence chains, and unified formatting. This not only aligns with industrial security auditing standards (e.g., OWASP Vulnerability Reporting Template, NVD CWE Documentation) but also directly reduces the "validation-to-fix" cycle for developers. As shown in Figure~\ref{fig:report}, the vulnerability reports generated now take the vulnerability lifecycle into full account, including root cause analysis, trigger conditions, potential impact assessment, and targeted fix recommendations—providing a comprehensive, end-to-end perspective that guides auditors and developers through every stage of vulnerability management, from identification to resolution.

\begin{tcolorbox}[colback=gray!20, colframe=gray!80, sharp corners, boxrule=0.5pt]
\textbf{Answer to RQ2:} FineSec improve vulnerability detection in three dimensions: \textbf{Accuracy improvement} - The LLaMA models stand out, achieving over a 20\% improvement, while Mistral delivers strong overall performance with a 73.3\% accuracy. \textbf{Deeper causes} -  LLMs delves deeper into the root causes and  focus more on the stability of the system’s long-term behavior as a chain risk. \textbf{Standardized reports} - LLMs generate structured reports containing CWE categories, vulnerability names, detailed explanations, and remediation suggestions, providing more practical reference value.
\end{tcolorbox}

\subsubsection{RQ3: How does the performance of LLMs vary across different CWE categorie?} To enable a fine-grained evaluation across vulnerability types, we classify the 30 selected CWEs into five categories based on their underlying causes and security implications, as shown in Table~\ref{tab:cwe_categories}. This classification is not arbitrary but grounded in the inherent similarities of the CWEs in terms of their underlying root causes and security implications, which allows for a more systematic analysis of model behavior.

\begin{table}[h]
  \caption{Common Weakness Enumeration (CWE) Vulnerability Categories}
  \label{tab:cwe_categories}
  \centering
  \begin{tabular}{>{\raggedright\arraybackslash}p{0.3\linewidth} >{\raggedright\arraybackslash}p{0.65\linewidth}}
    \toprule
    \textbf{Vulnerability Category} & \textbf{Associated CWE IDs} \\
    \midrule
    Memory Safety & 
    CWE-119, CWE-122, CWE-125, CWE-787, CWE-415, CWE-416, CWE-476 \\
    \addlinespace[0.2cm]
    Input Validation \& Injection &
    CWE-20, CWE-77, CWE-78, CWE-79, CWE-22, CWE-94, CWE-59, CWE-434 \\
    \addlinespace[0.2cm]
    System Resource \& Logic Errors &
    CWE-400, CWE-401, CWE-835, CWE-362, CWE-189, CWE-190 \\
    \addlinespace[0.2cm]
    Permissions \& Access Control &
    CWE-862, CWE-863, CWE-287, CWE-284, CWE-269, CWE-276 \\
    \addlinespace[0.2cm]
    Cryptography \& Information Leakage &
    CWE-295, CWE-310, CWE-200 \\
    \bottomrule
  \end{tabular}
\end{table}

Memory corruption vulnerabilities are among the most frequent potential problems. Dangling pointers, heap meta-data overwrites, uninitialized reads, and invalid or double-free vulnerabilities are common examples \cite{42}. Input Validation \&Injection does not rigorously validate or purify user input, allowing attackers to inject malicious codes or commands, resulting in data breaches, system takeover, or service interruptions. System Resource \& Logic Errors encompass three dimensions: resource lifecycle failures,currency violations, and boundary logic flaws. These vulnerabilities manifest non-deterministically under edge-case workloads, often leading to cascading failures. Permissions \& Access Control encompasses CWEs arising from flaws in access control mechanisms (e.g., missing permission checks, role confusion), with core security impacts including unauthorized operations or privilege escalation\cite{10.1145/3533703}. Cryptography \& Information Leakage encompasses CWEs stemming from improper use of cryptographic mechanisms (e.g., weak algorithms) or unintended exposure of sensitive information, directly threatening data confidentiality and integrity.

\begin{figure}[t]
\centering
\begin{minipage}{0.45\textwidth}
    \centering
    \includegraphics[width=0.9\linewidth, height=6cm]{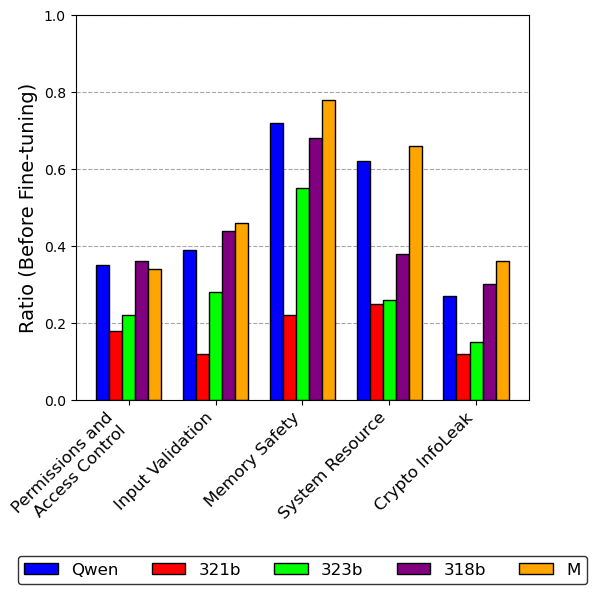}
    \caption*{{(a) Before FineSec}}
\end{minipage}
\begin{minipage}{0.45\textwidth}
    \centering
    \includegraphics[width=0.9\linewidth, height=6cm]{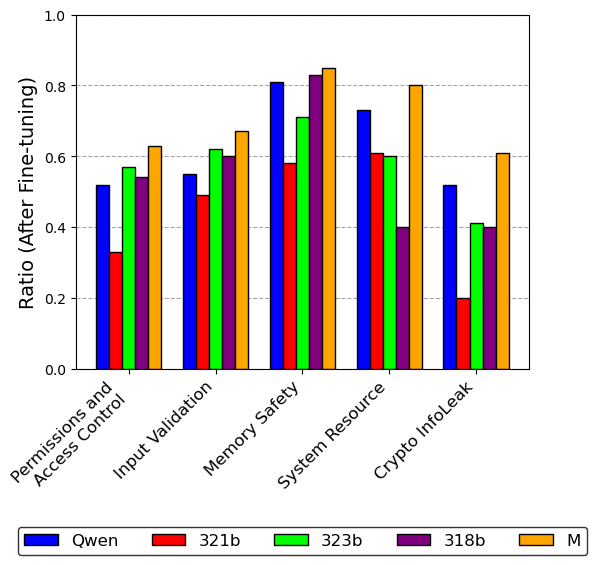}
    \caption*{{(b) After FineSec}}
\end{minipage}
\caption{Visual comparison demonstrating the effectiveness of FineSec in vulnerability detection and mitigation}
\label{fig:finesec_results}
\end{figure}

The experiment analyzed the model functionality through the CWE classification, revealing both the detection capabilities and the specific reasons behind the strengths and weaknesses of the models in different vulnerability categories. Figure~\ref{fig:finesec_results} presents a performance comparison across the CWE categories before and after FineSec processing.

The latest results reveal multidimensional performance variations across models: the 321B model demonstrates breakthroughs in previously weak foundational areas; the 323B model shows specialized expertise in access control; and the 318B model exhibits balanced optimization across categories. Qwen achieves remarkable progress in crypto security, while Mistral maintains comprehensive capabilities, building on its already high baseline performance.

In the Memory Safety domain, all models demonstrated strong detection capabilities, as these vulnerabilities are more explicit in code and require less contextual analysis. All models exhibited substantial improvements in their capabilities, achieving an average accuracy of 0.7448. Models with initially lower performance experienced significant gains after FineSec processing, while those with moderate baseline performance saw relatively smaller improvements. This discrepancy occurs because models with lower baseline performance had greater room for optimization and could more effectively absorb the targeted knowledge from the FineSec dataset, whereas models with moderate baseline performance were already closer to their performance ceiling for this specific task, leaving less margin for substantial improvement.

In System Resource \& Logic Erros category, llama series improved by around 0.3, gaining stronger concurrency detection and resource management capabilities. After fine-tuning, the Mistral model achieved a performance score of 0.789 in this category, making it particularly well-suited for high-concurrency service scenarios where it can efficiently detect related vulnerabilities and ensure stable system operation under heavy concurrent requests. In contrast, the 321B base model had an initial score of only 0.242, indicating a higher reliability risk in resource-intensive tasks.

In Permissions \& Access Control category, qwen and mistral showed significant improvement, suggesting that targeted fine-tuning for specific models can yield better results in permission-related vulnerability detection. After enhancement, the model demonstrated significantly heightened awareness of access control and permission-related risks. For instance, whereas it might have previously and indiscriminately categorized such issues under a broad label like "input validation," it now accurately identifies the underlying permission and access control—such as unauthorized privilege escalation, insecure direct object references, or broken authentication mechanisms—thereby providing more precise and actionable detection results.

For Input Validation \& Injection, all models exhibited consistent improvements. After enhancement, they were better able to identify unfiltered user inputs, validation logic bypasses, and other injection-related vulnerabilities, with increased detection accuracy for common exploitation scenarios.

Compared to the substantial improvements in the other four categories, cryptography and information leakage saw smaller gains, with all models improving by less than 0.258. This may be because cryptographic vulnerability detection often relies on complex cryptanalysis techniques (e.g., key entropy evaluation, side-channel attack detection), while the training data primarily covers basic encryption algorithms (e.g., Base64, MD5). As a result, models still struggle to identify more sophisticated real-world cryptographic vulnerabilities.

These findings suggest that current models possess reasonable detection capabilities for memory security vulnerabilities, but there is still scope for improvement in identifying more complex vulnerabilities. By analyzing the performance of models across specific CWE categories and comparing their respective advantages, we can gain a deeper understanding of the functionality of these tools and clarify future development directions. In particular, future research should prioritize enhancing the detection of complex vulnerabilities—especially those characterized by dynamic behavior and extensive system interactions.

\begin{tcolorbox}[colback=gray!20, colframe=gray!80, sharp corners, boxrule=0.5pt]
\textbf{Answer to RQ3:}Different models show distinct advantages in specific vulnerability detection. For instance, while LLaMA-32-1B has low overall accuracy, it performs well in memory management and system resource-related CWEs. After FineSec, LLaMA-32-3B excels in permissions and access control and LLaMA-31-8B is optimal for memory security vulnerabilities but needs optimization in encryption issues. Mistral has the strongest comprehensive capability and after fine-tuning, it also excels in Input Validation and has turned from a prior disadvantage in encryption-related detection to the top-performing model in this category. Qwen excels at addressing concurrency and other resource-related issues of the system.
\end{tcolorbox}

\subsubsection{RQ4:Does FineSec have the ability to detect new vulnerabilities?} We manually reviewed the vulnerability reports generated by FineSec on the test set and identified instances that, while not overlapping with the known ground truth labels, still exhibited clear security risks. These findings were further validated by security experts to ensure accuracy.

\begin{table}[t]
\caption{New Vulnerabilities Discovered by FineSec}
\label{tab:new_vulnerabilities}
\centering
\renewcommand{\arraystretch}{1.2}
\setlength{\tabcolsep}{0.6em}
\begin{tabular}{>{\raggedright\arraybackslash}p{2.8cm} 
                >{\raggedright\arraybackslash}p{6cm} 
                >{\raggedright\arraybackslash}p{4cm}}
\toprule
\textbf{Vulnerability Type} & \textbf{Description} & \textbf{Affected Function} \\
\midrule
XML/HTML5 Mixed Content Processing & 
Insufficient validation in HTML5/XML mixed context content processing, allowing unsafe content injection &
\texttt{unicode\_cp\_is\_allowed} \\

Improper Device Registration & 
Insecure BMC device release with unvalidated platform device registration and insecure dynamic device management &
\texttt{\_\_ipmi\_bmc\_register} \\

Debug vs Production Protection Mismatch & 
Reliance on debug-only checks instead of runtime protection, leading to missing validation in production environments &
\texttt{gup\_pte\_range} \\

Insecure Server Response Handling & 
Insecure server responses may lead to exposure of ACL data during transmission &
\texttt{nfs4\_call\_sync} \\

Hardware State Security Policy Asynchrony & 
Failure to verify terminal secure state before changing hardware status (baud rate, DTR line) or performing operations &
\texttt{do\_command} \\

Isolation Level Misuse & 
Using unverified isolation layer components (V8 engine's Isolate) directly may compromise isolation boundaries &
\texttt{callDebuggerMethod} \\

Cross-Process Window Management & 
Permission boundary violations and synchronization issues in cross-process window size manipulation &
\texttt{ShellWindowFrameView} \\

Code Structure Complexity & 
Overly complex code handling multiple execution paths creates opportunities for unauthorized access &
\texttt{tcp\_v6\_syn\_recv\_sock} \\

Insecure Transaction Cleanup & 
Transaction cleanup not properly associated with initialization status, resulting in resource leakage &
\texttt{zipfileBegin} \\
\bottomrule
\end{tabular}
\label{tab:new}
\end{table}

As shown in Table~\ref{tab:new}, FineSec successfully discovered nine previously undocumented vulnerabilities. This finding suggests that FineSec not only learns existing vulnerability patterns but also captures semantic and structural features indicative of previously unknown flaws.

\begin{tcolorbox}[colback=gray!20, colframe=gray!80, sharp corners, boxrule=0.5pt]
\textbf{Answer to RQ4:} FineSec has successfully detected nine previously unrecorded vulnerabilities in the C / C++ language that are not listed in the official CWE database. These results highlight FineSec’s potential as a security analysis tool capable of discovering vulnerabilities beyond existing classification systems. The insights from these new flaws could inform future updates to the CWE database, enhancing collective understanding of emerging vulnerability types and strengthening overall software security practices.  
\end{tcolorbox}

\section{Discussion}
\subsection{Related Work}
Recent years have seen notable progress in applying LLMs to C/C++ language vulnerability detection. Li et al. \cite{38} introduced IRIS, a framework that integrates LLMs with static analysis. By extracting CWE-specific taint specifications from third-party library APIs via LLMs and combining them with CodeQL, IRIS improves detection accuracy. To address false positives in static analysis, IRIS incorporates LLM-based context analysis, reducing developer workload while enhancing detection reliability. In the domain of dynamic testing, Huang et al. developed Fuzz4All\cite{10.1145/3597503.3639121}, which utilizes LLMs' understanding of language syntax to generate structurally valid yet semantically adversarial test cases. This approach enhances fuzzing efficiency by producing inputs that bypass superficial checks and target core program logic. 

Retrieval-Augmented Generation (RAG) techniques are employed to augment model capabilities in specific domains. For instance, Zhao et al. \cite{44} developed a curated knowledge base encompassing classic Python programming problems and their optimal solutions, which covers common logical operations involving data structures such as lists, strings, and dictionaries. This knowledge base serves as a retrieval resource to enhance the model's code generation accuracy by providing contextually relevant examples. Their subsequent work, Self Coder, introduces a self-guided framework that iteratively refines code drafts using a Code Interpreter, achieving state-of-the-art results on standard benchmarks. Wei et al. \cite{45} proposed SmartAuditFlow framework which dynamically customized audit strategies for smart contracts, combining structured reasoning with external tools to achieve high precision in vulnerability detection.They integrated iterative prompt optimization with external knowledge sources, including static analysis tools and RAG mechanisms. This integration ensures that audit decisions are grounded in contextual understanding and validated against real-world security knowledge, thereby enabling the production of comprehensive and authoritative security reports. Dozono et al. \cite{41} expanded vulnerability detection across five programming languages—C, Python, C++, Java, and JavaScript—and demonstrated GPT-4o’s superiority in both vulnerability detection and CWE classification, particularly under few-shot learning scenarios. Their results underscore the generalizability of LLMs but focus primarily on broad applicability rather than domain-specific optimization.

\subsection{Summary of Findings}Based on the above evaluation results, we have derived the following key findings:
\begin{itemize}

\item\textbf{Key Influencing Factors on Model Performance}: Through experiments, we found that the performance of the model is influenced by various factors such as its architecture, parameter size, and training data, as well as the uneven performance of LLM on different types of CWE. These findings reveal the inherent laws of model performance, providing insights for developing specialized models tailored to specific vulnerability categories and optimizing resource allocation in security analysis.

\item\textbf{Effectiveness of Knowledge Distillation and Fine-Tuning}: Extracting high-quality domain knowledge from powerful teacher models through multi-agent collaboration, and successfully transferring it to lighter student models through fine-tuning, FineSec significantly enhances the accuracy of models in detecting vulnerabilities. With proper fine-tuning, small models achieve or even surpass large commercial models. This underscores the potential of fine-tuned model as a valuable tool for improving vulnerability security.
% Our analysis also revealed a critical distinction between what we term "detectable" and "undetectable" vulnerabilities. Across all evaluated models, both our fine-tuned versions and the commercial APIs, performance was substantially higher on vulnerabilities characterized by clear, identifiable code patterns. I

\item\textbf{Resource Efficiency and Practical Deployment}: Our approach, particularly using QLoRA, allows effective model training and deployment on resource-constrained environments, such as a single NVIDIA Tesla T4 GPU with 16GB memory. This achievement showcases the practical feasibility of developing and deploying sophisticated LLM-based security solutions without the need for large-scale GPU clusters. 

\item\textbf{Framework Extensibility and Broader Applications}:
While this study was framed within the context of C/C++ vulnerability detection, the underlying framework is designed to be inherently modular and extensible. The core principles of our methodology, knowledge distillation from expert models and the teacher-student collaborative architecture, are not specific to any single programming language or application domain. This inherent flexibility allows the FineSec framework to be readily adapted to other critical areas. 

\end{itemize}

\subsection{Threats of Validity} Although our proposed system demonstrates promising results in enhancing LLM capabilities for C/C++ language vulnerability detection, several potential limitations must be acknowledged. These limitations highlight factors that may influence the generalizability and robustness of our findings.

\begin{itemize}
\item Input length constraints:
Due to the inherent token-length limitations of current LLM architectures, our experiments use code snippets as inputs rather than full project files. While snippet-level evaluation provides valuable insights into detection capabilities, it does not fully capture long-range dependencies, cross-file interactions, or system-level vulnerabilities that are common in real-world software systems. This constraint may limit the applicability of our findings to scenarios involving large code bases or highly interdependent modules. Future work could integrate hierarchical modeling or retrieval-augmented generation (RAG) to incorporate broader program context while respecting token constraints.

\item Limited performance in context-dependent and logically complex scenarios:
Our evaluation indicates that the model performs less effectively in detecting vulnerabilities requiring deep contextual reasoning or multi-step logical inference, such as concurrency issues (e.g., race conditions, deadlocks) and cryptographic weaknesses (e.g., flawed key management, side-channel leaks). These categories often demand semantic reasoning across multiple code paths or modules, which static snippet-based analysis cannot fully address. 
% To improve performance, future research could leverage teacher–student training with specialized expert models for complex vulnerability domains, incorporate Abstract Syntax Trees (ASTs) and control/data-flow graphs generated from static analysis tools to enrich the model’s structural understanding, and employ multi-agent collaboration between static and dynamic analysis pipelines to integrate such structural insights with runtime behavior analysis.
\item Domain transferability:
While our study focuses on the C/C++ programming language, many real-world systems are polyglot environments where vulnerabilities may arise from cross-language interactions (e.g., C / C++ with Python bindings, C with WebAssembly). The current system’s performance in such heterogeneous environments remains untested. Future work should evaluate FineSec in cross-language contexts and adapt the methodology to additional programming languages.
\end{itemize}
% In summary, these limitations suggest that while our system significantly improves LLM-based vulnerability detection, further research is required to improve its scalability, adaptability, and performance in complex real-world scenarios. Addressing these threats will be the key to transitioning FineSec from a controlled research prototype to a deployable practical security analysis tool.

\section{Conclusion}

In this paper, we conducted extensive benchmark experiments to evaluate the capabilities of LLMs in detecting vulnerabilities in C/C++ programs. Our findings revealed both the promising potential of LLMs and their current limitations, motivating the development of FineSec—a comprehensive framework designed to transform general-purpose LLMs into domain-specific security experts. FineSec follows a four-stage pipeline encompassing data preparation, model training, performance evaluation, and continuous learning. It leverages advanced techniques such as knowledge distillation, parameter-efficient fine-tuning, and instruction alignment to enhance detection accuracy and improve adaptability across diverse vulnerability categories.

Experimental results demonstrate that FineSec consistently improves detection performance across multiple baseline models. Remarkably, some small-parameter models, after FineSec optimization, outperformed larger models in real-world vulnerability detection tasks, offering high-performance security analysis without the prohibitive computational costs typically associated with large-scale models. Beyond vulnerability identification, FineSec-powered models also generate actionable repair suggestions, increasing their practical value in secure software development workflows. From a vulnerability category perspective, FineSec achieved substantial gains in memory safety and input validation, while highlighting the need for further enhancements in concurrency-related issues and cryptographic vulnerabilities. These insights provide a roadmap for targeted improvements in areas where current LLM capabilities remain limited. Future work will focus on enriching complex vulnerability detection by incorporating semantic representations such as abstract syntax trees (ASTs) and program dependency graphs (PDGs) into the training process. We also aim to strengthen repair recommendations through automated patch synthesis combined with explanatory reasoning, enabling developers to understand both the fix and its underlying rationale. Furthermore, we plan to extend FineSec’s applicability beyond the C/C++ language to a wider range of programming languages and security-critical domains, including smart contract auditing and embedded system firmware analysis.

%Bibliography
\bibliographystyle{unsrt}  
\bibliography{references}

\end{document}